\documentclass{article}
\usepackage{arxiv}
\usepackage[utf8]{inputenc} 
\usepackage[T1]{fontenc}    
\usepackage{hyperref}       
\usepackage{url}            
\usepackage{booktabs}       
\usepackage{amsfonts}       
\usepackage{nicefrac}       
\usepackage{microtype}      
\usepackage{graphicx}
\usepackage{natbib}
\usepackage{doi}
\usepackage[labelfont=bf]{caption}
\usepackage{amsmath,amsfonts,amssymb,amsthm,booktabs,color,epsfig,graphicx,hyperref,url}
\usepackage{bm}
\usepackage{multirow}
\theoremstyle{plain}
\newtheorem{theorem}{Theorem}

\newtheorem{proposition}{Proposition}

\newtheorem{corollary}{Corollary}
\theoremstyle{definition}

\newcommand\BibTeX{{\rmfamily B\kern-.05em \textsc{i\kern-.025em b}\kern-.08em
T\kern-.1667em\lower.7ex\hbox{E}\kern-.125emX}}
\bibpunct{(}{)}{;}{a}{,}{,}  

\title{On the distribution of isometric log-ratio transformation
under extra-multinomial count data}
\date{} 			

\author{ \href{https://orcid.org/0000-0001-7859-0483}{\includegraphics[scale=0.06]{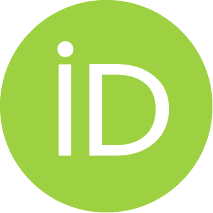}\hspace{1mm}Noora Kartiosuo}$^{1,2,3,4}$, 
 \href{https://orcid.org/0000-0002-2150-2769}{\includegraphics[scale=0.06]{orcid.pdf}\hspace{1mm}Joni Virta}$^{1}$, 
\href{https://orcid.org/0000-0001-6295-0245}{\includegraphics[scale=0.06]{orcid.pdf}\hspace{1mm}Jaakko Nevalainen}$^{5}$,
\href{https://orcid.org/0000-0001-9365-3702}{\includegraphics[scale=0.06]{orcid.pdf}\hspace{1mm}
Olli Raitakari}$^{2,3,6}$,  
\href{https://orcid.org/0000-0002-8162-2260}{\includegraphics[scale=0.06]{orcid.pdf}\hspace{1mm}Kari Auranen}$^{1,7}$\\
}
\renewcommand{\headeright}{ }
\renewcommand{\undertitle}{ }
\renewcommand{\shorttitle}{Distribution of ilr coordinates}

\begin{document}

\maketitle
\noindent {1 Department of Mathematics and Statistics, University of Turku, Turku, Finland \\
2 Research Centre of Applied and Preventive Cardiovascular Medicine, University of Turku, Turku, Finland \\
3 Centre for Population Health Research, University of Turku, Turku, Finland  \\
4 Murdoch Children's Research Institute, Melbourne, Australia \\ 
5 Health Sciences Unit, Faculty of Social Sciences, Tampere University, Tampere, Finland\\
6 Department of Clinical Physiology and Nuclear Medicine, University of Turku and Turku University Hospital, Turku, Finland\\
7 Department of Clinical Medicine, University of Turku, Turku, Finland\\
 } \\

\section*{Acknowledgments}
 NK has been financially supported by Emil Aaltonen Foundation and the MATTI programme in The University of Turku Graduate School (UTUGS). The work of JV was supported by Research Council of Finland, Grants 347501 and 353769.
 \\

\begin{abstract}
Compositional data can be mapped from the simplex to the Euclidean space through the isometric log-ratio (ilr) transformation.  When the underlying counts follow a multinomial distribution, the distribution of the ensuing ilr coordinates has been shown to be asymptotically multivariate normal. We  derive conditions under which the normality of the ilr coordinates holds under a compound multinomial distribution inducing overdispersion in the counts. We investigate the practical applicability of the approximation under extra-multinomial variation using a simulation study under the Dirichlet-multinomial distribution. The approximation works well, except with a small total count or high amount of overdispersion. Our work is motivated by microbiome data, which exhibit extra-multinomial variation and are increasingly treated as compositions. We conclude that if empirical data analysis relies on the normality of ilr coordinates, it may be advisable to choose a taxonomic level with less sparsity so that the distribution of taxon-specific class probabilities remains unimodal.
\end{abstract}


\section{Introduction\label{sec:1}}
Compositional data involve elements which together constitute a whole entity. Such data may arise when non-negative observations, such as counts, are considered as compositions through scaling individual observations by their total into proportions that sum up to unity. The proportions lie in the simplex, and due to the unit-sum constraint, traditional statistical methods may not be applicable. For these type of data, the framework of compositional data analysis provides a suitable collection of methods \citep{pawlowskyglahn_compositional_2011}.

Importantly, the simplex has a Euclidean vector space structure where different coordinate systems can be specified using different log-ratio transformations \citep{Mateu-Figuerasbook3}. Particularly convenient is the isometric log-ratio transformation (ilr), which maps the simplex  equipped with the so called Aitchison geometry to the standard Euclidean vector space of coordinates within an orthonormal basis \citep{Egozcue2003}. The isometric structure of the two spaces allows basing probabilistic modelling for vectors within the simplex directly on their ilr coordinates  \citep{Mateu-Figuerasbook3}.  In particular, the so called normal on the simplex is obtained by assuming that the ilr coordinates have a standard multivariate normal distribution \citep{Mateu-Figueras2013}. The benefit of “working on coordinates” is that standard statistical methods, such as linear regression models, are readily applicable.

Recently, the use of compositional data analysis has been endorsed also in microbiome research, where datasets consist of read counts of a number of distinct taxa, but the total of the counts, resulting from sequencing depth,  does not carry any substantive information, being arbitrary  \citep{Gloor2017}. Different log-ratio transformations have been applied to microbial counts assuming the normality of the log-ratio coordinates \citep{Sohn2019, Zhang2019a}. When the underlying counts follow the multinomial distribution, the induced distribution of the ensuing ilr coordinates, viewed basically as a data transformation, has indeed been shown to be asymptotically multivariate normal \citep{Graffelman}. Thus, given a large enough total count, it is justifiable to apply methods based on the assumption of normality of ilr coordinates. Furthermore, the mean and covariance parameters of the limiting distribution of the ilr coordinates can then be expressed in terms of the multinomial probability parameters \citep{Graffelman}.

Nevertheless, microbial counts are characterised by heterogeneity across individuals in taxon-specific class probabilities, which leads to overdispersion of class-specific counts in regard to the simple multinomial distribution. Such heterogeneity can be addressed by treating counts as realisations from a compound multinomial distribution where the class-specific probabilities are not fixed but follow some distribution. A possible choice is the logratio-normal-multinomial distribution, where the class-specific proportions are modelled as inverses of the ilr coordinates. \citep{Mateu-Figueras2013} However, even this approach assumes that the ilr coordinates are normally distributed.
Alternatively, over-dispersed counts are obtained when the class probabilities follow some distribution. The standard choice among such compound multinomial distributions is the Dirichlet-multinomial model. Although criticised due to the negative correlations between parts of compositions \citep{Xia2013, ComasCufi2020}, 
this distribution has often been used in the analysis of microbial  data. \citep{Fernandes2013, Wang2019}

At the extreme levels of heterogeneity, the distribution of proportions becomes multimodal, causing sample-specific counts to concentrate within one of the classes and leading to zero counts in many other classes. 
It is obvious that under such a large overdispersion  the distribution of class-specific proportions is multimodal and the asymptotic normality of the induced distribution of the ilr coordinates cannot hold.
However, it still remains to be investigated how the amounts of extra-multinomial overdispersion affect the normality of the ilr coordinates.  The aim of this paper is to investigate conditions under which the asymptotic normality of the ilr coordinates holds when  based on compound multinomial counts, i.e., on count observations that exhibit extra-multinomial variation. Apart from a general theoretical result we consider a simulation study where, for convenience and transparency of interpretation, we focus on counts that follow a Dirichlet-multinomial distribution.

The contributions of this work are as follows:
\begin{itemize}
\item We present conditions under which the asymptotic normality of  compound multinomial counts holds (Section 2).
\item Assuming a Dirichlet-multinomial distribution of counts, we present conditions for the  asymptotic normality of counts and the ensuing  proportions (Section 3). 
\item Based on the asymptotic normality of the proportions under compound multinomial models, we 
argue for the asymptotic normality of the induced ilr coordinates and derive an explicit normal approximation under the special case of the Dirichlet-multinomial model (Section 4).
\item Using a simulation study, we investigate how well the approximation under the Dirichlet-multinomial holds  in the presence of varying extents of extra-multinomial variation (Section 5).
\item Additionally, we study how  variability in the  sample-specific total count influences the distribution of the ilr coordinates (Sections 4 and 5).
\end{itemize}

\section{Asymptotic normality of compound multinomial distributions}\label{general} 
 Denote the $J$-dimensional closed unit simplex by $\mathcal{S}^{J - 1}$. We here consider the asymptotic behaviour of a compound multinomial variable $X_K$, $X_K\mid \Pi_K \sim\mathrm{Multinomial}(K,\Pi_K)$  where the probability parameter $\Pi_k \in S^{J-1}$ admits a decomposition 
 $\Pi_K = \alpha+ Z_K$, where $\alpha\in S^{J-1}$ is fixed and $Z_k$  is a
random  $J$-vector. The distribution of $Z_K$ thus determines the mixing distribution for the vector of multinomial probabilities. The index $K$ corresponds to the total count of the observation $X_K$ and we present our results in the asymptotic scenario where $K \rightarrow \infty$. This regime approximates the practical scenario where the total observed count is sufficiently large. In this section, we state two results concerning the  asymptotic normality of $X_K$ under two alternative conditions for the mixing distribution. The formal proofs are provided in the Appendix. Then, in Section 3, we apply these results to derive an asymptotic normal approximation for $X_K$ in the special case where the mixing distribution is Dirichlet. Note that throughout this paper we consider the asymptotic distributions of {\sl observations} (counts and their proportions) and the induced ilr coordinates, rather than the asymptotic behaviour of parameter estimators.

The two theorems are based on alternative conditions, respectively. The first condition requires fast enough convergence of $||Z_K||^2$ to 0 as $K\to\infty$.
The  convergence of $||Z_K||^2$ then warrants that the role of the mixing distribution is asymptotically negligible and the asymptotic normality of $X_K$ follows from the normal limiting distribution of the multinomial distribution (Theorem 1).
The second condition assumes the existence of a limiting distribution for an appropriately scaled $Z_K$, as $K\to \infty $. Here, the mixing distribution is assumed to converge to its limiting distribution slower than the multinomial distribution and thus the limiting distribution becomes that of the mixing distribution (Theorem 2).

\begin{theorem}\label{theorem_1}
We consider a sequence of $J$-vectors $Z_K =  (Z_{K1}, \ldots, Z_{KJ})'$.  
	Let $\| \cdot \|$ denote the Euclidean norm and assume that
	\begin{align*}
	K \mathrm{E} ( \|Z_K\|^2 ) &= o(1)
	\end{align*} 
	when $K \rightarrow \infty$. 
  Then
	\begin{align*}
	\frac{1}{\sqrt{K}}(X_K - K {\bm\alpha}) \rightsquigarrow \mathcal{N}(0, \hbox{diag}(\bm{\alpha}) - {\bm\alpha} {\bm\alpha}'),
	\end{align*}
 where $\rightsquigarrow$ denotes convergence in distribution.
\end{theorem}

\begin{theorem}\label{theorem_2}
	Assume that
	\begin{align*}
	\sqrt{\alpha_K} Z_K \rightsquigarrow \mathcal{D},
	\end{align*} 
	for some rate $\alpha_K$ and distribution $\mathcal{D}$, when $K \rightarrow \infty$. Assume further that $\alpha_K/K = o(1)$. Then
	\begin{align*}
	\frac{\sqrt{\alpha_K}}{K}(X_K - K {\bm\alpha}) \rightsquigarrow \mathcal{D}.
	\end{align*}
\end{theorem}

\section{The special case of the Dirichlet-multinomial distribution}\label{diri}
We apply Theorems 1 and 2 in the special case in which the mixing distribution is Dirichlet, i.e., when the class-specific probabilities of a multinomial model vary according to a Dirichlet distribution. 
We thus consider the following hierarchical model:
\begin{align*} 
&\Pi_K = (\Pi_{K1}, \ldots, \Pi_{KJ}) \sim \hbox{Dirichlet}(\alpha_{K1}, \ldots, \alpha_{KJ}), \\
&X_K  \vert \Pi_K = (X_{K1},\ldots,X_{KJ})  \sim \hbox{Multinomial}(K, \Pi_K).\nonumber
\end{align*}
We reparametrise the Dirichlet distribution
in terms of $\tilde\alpha_j = \alpha_{Kj}/\alpha_K$, $j=1, \ldots, J$, where $\alpha_K=\sum_{j=1}^J\alpha_{Kj}$, making the simplifying assumption that the $\tilde\alpha_j$'s are independent of $K$. 
Parameter $\alpha_K$ controls the heterogeneity of class probabilities as compared to the purely multinomial distribution. 
We call  $\alpha_K$ sparsity,  because the smaller its value, the more  heterogeneous the class probabilities become between different samples, i.e., the more sparse the ensuing counts are. Note that we consider $\alpha_K$ as a sequence of values indexed by $K$ and with the limit $\alpha_K\to\infty$ as $K\to\infty$. We  assume that each $\tilde\alpha_j\alpha_K >1$, and hence $\alpha_K > J$, so that the Dirichlet distribution has a unique mode $((\alpha_{K1}, \ldots, \alpha_{KJ}) - 1_J)/(\alpha_K- J)$. Here $1_J$ is the $J$-vector $(1,\ldots,1)^T$.
  We denote 
  the vector  $(\tilde\alpha_1, \ldots,\tilde\alpha_J)^T$ by $\tilde{\boldsymbol{\alpha}}$. 
 We delineate two cases according to whether $K = o(\alpha_K)$, i.e., $K/\alpha_K \to 0$, or $\alpha_K = o(K)$, i.e., $\alpha_K/K\to 0$, when both $\alpha_K\to\infty$ and $K\to\infty$.

\begin{corollary}
	Let $X_K \sim \mathrm{Multinomial}(K, \Pi_K)$ and  $\Pi_K \sim \mathrm{Dirichlet}(\alpha_K\tilde{\bm\alpha})$ where $\tilde{\bm\alpha} \in \mathcal{S}^{J - 1}$. Then, the following hold. If $K/\alpha_K = o(1)$, then
		\begin{align*}
		\frac{1}{\sqrt{K}}(X_K - K \tilde{\bm\alpha}) \rightsquigarrow \mathcal{N}(0, \hbox{diag}(\boldsymbol{\tilde\alpha})- \tilde{\bm\alpha} \tilde{\bm\alpha}').
		\end{align*}
\end{corollary}

Let $p_{Kj} = X_{Kj}/K$, $j=1, \ldots, J$, denote the proportions (i.e. relative frequencies) based on Dirichlet-multinomial counts $X_K$. Denote 
$\textbf{p} =  (p_{K1}, \ldots, p_{KJ})^T$. Under the conditions for Corollary 1, it follows immediately  that 
\[
\sqrt{K}(\textbf{p} - \tilde{\bm\alpha})  \rightsquigarrow \mathcal{N}(0,\boldsymbol{\Sigma_\Pi}),
\]
where $\boldsymbol{\Sigma_\Pi} = 
\hbox{diag}(\boldsymbol{\tilde\alpha}) - \tilde{\bm\alpha} \tilde{\bm\alpha}'$. 
With a large enough total count $K$ and $\alpha_K$ the distribution of the proportions can thus be viewed as normal and we write
\begin{equation}\label{pjak_cor1}
\textbf{p} \simeq \mathcal{N}(\boldsymbol{\tilde{\alpha}}, \frac{1}{K}\boldsymbol{\Sigma_\Pi}).
\end{equation}

\begin{corollary}
		If $\alpha_K/K = o(1)$, then
		\begin{align*}
		\frac{\sqrt{\alpha_K+1}}{K}(X_K- K \tilde{\bm\alpha}) \rightsquigarrow \mathcal{N}(0, \hbox{diag}(\boldsymbol{\tilde\alpha}) - \tilde{\bm\alpha} \tilde{\bm\alpha}').
		\end{align*}
\end{corollary}
Corollary 2 above follows from the fact that under the assumption of unimodality, the sequence $\Pi_K$ has a limiting normal distribution (for details, see the Appendix). Interestingly, the limiting distributions in Corollary 1 and Corollary 2 are the same. The difference between the two scenarios is only in the rate of convergence.  

It follows from Corollary 2 that for large enough values of $K$ and $\alpha_K$ we can write 
\begin{equation}\label{pjak_cor2}
\textbf{p} \simeq \mathcal{N}(\boldsymbol{\tilde{\alpha}}, \left(\frac{1}{\alpha_K +1}\right)\boldsymbol{\Sigma_\Pi}).
\end{equation}

The asymptotic normality of the proportions holds when both  $K\to\infty$ and $\alpha_K\to\infty$, regardless of which of the two parameters grows faster. As shown above, when $K \gg \alpha_K$, the asymptotic behaviour of the distribution of the proportions is dominated by the asymptotic normality of the multinomial distribution. By contrast, when $\alpha_K \gg K$,  the behaviour is dominated by the asymptotic normality of the mixing distribution (Dirichlet).

Results (\ref{pjak_cor1}) and (\ref{pjak_cor2}) can be combined by writing 
\begin{equation}\label{p_as_jak}
\textbf{p} \simeq \mathcal{N}(\boldsymbol{\tilde{\alpha}},\boldsymbol{\Sigma_p}),
\end{equation}
where 
\begin{equation}\label{sigmap}
\boldsymbol{\Sigma_p} = \frac{1}{K}\left( \frac{\alpha_K + K}{\alpha_K +1}\right)\boldsymbol{\Sigma_\Pi}.
\end{equation}
In particular, $\boldsymbol{\Sigma_p}$ reduces to those in expressions (\ref{pjak_cor1}) or (\ref{pjak_cor2}) when
 $\alpha_K \gg K $
or $K \gg \alpha_K$, respectively. The formulation above is 
motivated by the fact that $\boldsymbol{\Sigma_p}$ is the variance-covariance matrix of relative frequencies of Dirichlet-multinomial counts, as found easily by the law to total variance.
 Of note, the excess variability of proportions \textbf{p} under the Dirichlet-multinomial distribution as compared to those under the multinomial distribution is 
\begin{equation}\label{excessvar_Dirmn}
(\alpha_K + K)/(\alpha_K +1). 
\end{equation}

\subsection{Overdispersed  total count}
Next, we extend the Dirichlet-multinomial model to allow ovedspersion not only in the proportions across the samples, but also in the total count. In general, empirical data on read counts often exhibit both types of overdispersion. 
A typical model for the total count is the negative binomial distribution (i.e. overdispersed Poisson). However, we here formulate the model of the total count using the log-normal distribution for it will allow an explicit expression of the moments of the marginal distribution of the proportions.  In practice, while the log-normal distribution is continuous, the values for the total count are simply  rounded to the nearest integer. 

We denote the distribution as $K \sim $ lognormal($\mu$,$\sigma^2$), with the expectation and variance
\begin{align}\label{lognormaljak}
&\hbox{E}(K) = \hbox{exp}(\mu+  \sigma^2/2)   \\ \nonumber
&\hbox{Var}(K) =(\hbox{exp}({\sigma^2})- 1)\hbox{exp}(2\mu+ \sigma^2).
\end{align}

It follows from (\ref{p_as_jak}) and (\ref{lognormaljak}) that the marginal expectation of the vector of proportions \textbf{p} is 
\begin{align*}
\hbox{E}( \textbf{p} )  = \hbox{E}_K \left( \hbox{E} \left(  \textbf{p}  | K \right) \right) = \hbox{E}_K (\tilde{\boldsymbol{\alpha}}) =  \boldsymbol{\tilde{\alpha}}. 
\end{align*} 
Similarly, by the law of total covariance, the variance-covariance matrix of  $\textbf{p}  $ is 
\begin{align}\label{varp_lndirmn}
&\boldsymbol{\Sigma_p} = \hbox{E}_K \left( \hbox{Cov}(\textbf{p}, \textbf{p}^T| K) \right) + \hbox{Cov}_K \left(\hbox{E}\left(\textbf{p} |K \right) , \hbox{E}\left( \textbf{p}^T | K\right)  \right) \nonumber \\
&=\hbox{E}_K \left( \frac{1}{K}  \left(\frac{\alpha_K + K}{\alpha_K +1} \right) \boldsymbol{\Sigma}_{\boldsymbol\Pi} \right) 
+ \hbox{Cov}_K \left( \boldsymbol{\tilde\alpha} \boldsymbol{\tilde\alpha}^T  \right) \\
&= \boldsymbol{\Sigma}_{\boldsymbol\Pi}\left( \frac{1}{\alpha_K+1}\right)\left(\alpha_K \hbox{exp}(-\mu + \sigma^2/2)+1
\right) \nonumber = \boldsymbol{\Sigma}_{\boldsymbol\Pi}\gamma(\alpha_K, \mu, \sigma^2), \nonumber
\end{align}
where we denote
\begin{align}\label{dir_ln_mn_exvar}
\gamma(\alpha_K, \mu, \sigma^2) = \left( \frac{1}{\alpha_K+1}\right)
\left(\alpha_K \hbox{exp}(-\mu + \sigma^2/2)+1
\right).
\end{align}

\begin{proposition} 
When the distribution of the total count is lognormal($\mu$, $\sigma^2$),
a normal approximation for the distribution of the proportions \textbf{p} is given by
\begin{equation*}
\textbf{p}  \approx \mathcal{N}(\tilde{\boldsymbol{\alpha}}, \gamma(\alpha_K, \mu, \sigma^2) \boldsymbol{\Sigma}_{\boldsymbol\Pi}).
\end{equation*}

\end{proposition}

We here have used the property of the lognormal distribution according to which the distribution of $1/K$ is lognormal($-\mu$, $\sigma^2$) if the distribution of $K$ is lognormal($\mu$, $\sigma^2$).  The excess variability of proportions under  lognormal total count, incomparison to the multinomial distribution with total count $K$ chosen to correspond to the median of the lognormal distribution, is $\hbox{exp}(\mu) \gamma(\alpha_K,\mu, \sigma^2)$. In the limit $\alpha_K \rightarrow \infty$, this ratio is $\hbox{exp}(\mu)\hbox{exp}(-\mu + \sigma^2/2) = \hbox{exp}(\sigma^2/2)$. We investigate the performance of the proposed approximation in Section\ref{rare}.

\section{Ilr coordinates and their asymptotic normality}
In this section we first review the isometric log-ratio transformation when  applied to compositional count data to transform proportions within a simplex to Euclidean coordinates. 
We then argue about the asymptotic normality of the ilr coordinates when based on compound multinomial counts under the conditions of Theorems 1 and 2. We derive an explicit normal approximation of the ilr coordinates under the Dirichlet-multinomial model.  We also extend our approximation to scenarios with varying total count.

\subsection{Contrasts and the ilr coordinates}
Proportions based on counts are here considered as  a composition, constrained by the unit-sum condition  $\sum_{j=1}^J p_{j} = 1$. To transform  proportions from the $J$-dimensional simplex to  $(J-1)$-dimensional Euclidean coordinates, we apply the isometric log-ratio (ilr) transformation \citep{Egozcue2003}.  

The first step is to define an appropriate set of contrasts between the $J$ components using a sequential binary partition (SBP). There is no canonical basis for choosing the contrasts \citep{pawlowskyglahn_compositional_2011}. They can be based, for example,  on  known relationships between the parts of the composition. A common way to define a partition is \textit{pivotal}, where one part is contrasted against all other classes, and the other parts of the composition are sequentially contrasted against the remaining parts. This leads to the following SBP matrix: 
\begin{equation}\label{pivotalmat}
\boldsymbol{\Psi} = (\psi_{jk})=
\begin{bmatrix}
+1 &  0  & 0 & \ldots & 0 \\
-1 & +1 & 0 & \ldots & 0 \\
-1 & -1 & +1 & \ldots & 0 \\
\vdots & \vdots & \vdots  & \ddots & \vdots  \\
-1 & -1 & -1 & \vdots & +1 \\
-1 & -1 & -1 & \vdots & -1 \\
\end{bmatrix}.
\end{equation}

Let \textbf{M} denote the vector of ilr coordinates. 
With any given  SBP  matrix $\boldsymbol{\Psi}$, the ilr transformation is  obtained  as follows  \citep{Graffelman}:
\begin{align*}
\textbf{M}= \hbox{ilr}(\textbf{p}) = \textbf{V}^T \hbox{ln}(\textbf{p)}, 
\end{align*}
where the contrast matrix \textbf{V}  is based on $\boldsymbol{\Psi}$: 
\begin{align*}
(\textbf{V})_{jk} = \psi_{jk}\sqrt{\frac{n^+_k n^-_k}{n^+_k + n^-_k}} 
\left[\frac{1}{n_k^+}\right]^{\textbf{I}[\textbf{sign}(\psi_{jk})=+]}
\left[\frac{1}{n_k^-}\right]^{\textbf{I}[\textbf{sign}(\psi_{jk})=-]}. 
\end{align*}
Here,  $n_k^+$ and $n_k^-$ are the numbers of cells with values $+1$ and $-1$ in the $k$th column of $\boldsymbol{\Psi}$.  
In general, the above definition means that ilr coordinates can be calculated as
\begin{align*}
M_{k} = \sqrt{\frac{n^+_k n^-_k}{n^+_k + n^-_k}}  \hbox{ln} \frac{\hbox{gm}(p_{k}^+)}{\hbox{gm}(p_{k}^-)}, k=1, \ldots, J-1,
\end{align*}
where $\hbox{gm}(p_{k}^+)$ denotes the geometric mean of the proportions in the classes denoted by $+1$. Whereas the original proportions lie in the simplex, the transformed coordinates reside in the  Euclidean space.

\subsection{Asymptotic distribution of the ilr coordinates}\label{as_norm}
The asymptotic normality of the ilr coordinates when based on the (ilr) transformation of purely multinomial counts has been shown earlier \citep{Graffelman}.  Based on our  Theorems 1 and 2,  it  follows immediately by the delta method that the asymptotic normality of the ilr coordinates holds under much general conditions, allowing extra-multinomial variation.  To demonstrate this, we here derive an explicit asymptotic normal approximation to the ilr coordinates under the special case of Dirichlet-multinomial counts.

Based on the delta method, a transformation $g$ of  proportions \textbf{p} based on the Dirichlet-multinomial mixture satisfies asymptotically 
\begin{equation*}
g(\textbf{p})  \approx \mathcal{N}\left( g(\hbox{E}(\textbf{p})),
\left(\frac{\partial g(\boldsymbol{\textbf{p}})}{\partial \boldsymbol{\textbf{p}}}\right) \boldsymbol{\Sigma}_{\boldsymbol{p}} \left(\frac{\partial g(\boldsymbol{\textbf{p}})}{\partial \boldsymbol{\textbf{p}}}\right)^T
\right),
 \end{equation*} 
where $\boldsymbol{\Sigma}_{\textbf{p}}$ is the variance-covariance matrix  (\ref{sigmap}) and the derivatives are evaluated at $\hbox{E}(\textbf{p}) = \boldsymbol{\tilde\alpha}$.  
For $g(\textbf{p}) = \hbox{ilr}(\textbf{p})$,  the derivatives  are
\begin{equation*}
\frac{\partial \hbox{ilr} ( \boldsymbol{\textbf{p}})}{\partial \boldsymbol{\textbf{p}}}  =
\frac{\partial \textbf{V}^T\hbox{ln}(\boldsymbol{\textbf{p}})}{\partial \boldsymbol{\textbf{p}}} = 
\textbf{V}^T \textbf{D}^{-1}_{\boldsymbol{\tilde\alpha}}.
\end{equation*}

Based on the delta method, we thus arrive at the asymptotic distribution for the ilr coordinates.
\begin{corollary}
The  asymptotic distribution for the ilr coordinates is obtained as follows: 
\begin{equation}\label{ilrjakauma_noncent}
\hbox{ilr}(\textbf{p}) \approx \mathcal{N} \left(\hbox{ilr}(\boldsymbol{\tilde{\alpha}}),  
\textbf{V}^T \textbf{D}^{-1}_{\boldsymbol{\tilde{\alpha}}}\boldsymbol{\Sigma_p} \textbf{D}^{-1}_{\boldsymbol{\tilde{\alpha}}} \textbf{V} \right).
\end{equation}
Under multinomial counts, the variance-covariance matrix is simplified to $
(1/K)\textbf{V}^T \textbf{D}^{-1}_{\boldsymbol{\tilde{\alpha}}}\textbf{V} 
$  \citep{Graffelman}. 
In case of Dirichlet-multinomial counts, we can similarly write  the variance-covariance matrix as $(1/K)(\alpha_K + K)/(\alpha_K +1)  \textbf{V}^T \textbf{D}^{-1}_{\boldsymbol{\tilde{\alpha}}}\textbf{V}$.
\end{corollary}

In practice, ilr($\boldsymbol{\tilde\alpha}$) may deviate considerably from E(ilr(\textbf{p})).
For any scalar random variable $x$, the expectation of $\hbox{ln}(x)$ is approximated bay 
\[\hbox{E}(\hbox{ln}(x)) \simeq  \hbox{ln}(\hbox{E}(x)) - \frac{\hbox{Var}(x)}{2\hbox{E}(x)^2},\]
which holds well for large $x$.
Applying the above approximation separately for each element of \textbf{p}, we obtain 
\[
\hbox{E}(\hbox{ln}(\textbf{p}))  \simeq \hbox{ln}(\boldsymbol{\tilde{\alpha}}) - (1/2)(1/K)((\alpha_K + K)/(K\alpha_K + K)) \ \bm\lambda \ \circ \ \boldsymbol{\beta},
\]
where $\boldsymbol\beta = (1/\tilde\alpha_1^2, \ldots, 1/\tilde\alpha_J^2)$ and $\bm\lambda$ is a vector of the diagonal elements of $\boldsymbol{\Sigma}_{\boldsymbol\Pi}$.
Using this alternative expression, the asymptotic distribution of ilr coordinates based on Dirichlet-multinomial counts is 
\begin{equation}\label{ilrjakauma_alt}
\hbox{ilr}(\textbf{p}) \approx \mathcal{N} \left(\textbf{V}^T (\hbox{ln}(\boldsymbol{\tilde{\alpha}}) - (1/2)(1/K)((\alpha_K + K)/(K\alpha_K + K)) \ \bm{\lambda} \ \circ \ \boldsymbol{\beta}),   
(1/K)(\alpha_K + K)/(\alpha_K +1)  \textbf{V}^T \textbf{D}^{-1}_{\boldsymbol{\tilde{\alpha}}}\textbf{V} \right).
\end{equation}

 Finally, if the total count follows the log-normal distribution instead of being fixed, the variance-covariance matrix $\boldsymbol{\tilde\Sigma_{p}}$  of the proportions is given by (\ref{varp_lndirmn}) and thus the covariance matrix in (\ref{ilrjakauma_alt}) depends on  parameters $\mu$, $\alpha_K $ and $\sigma^2$. 
  Thus, the distribution of the ilr coordinates is now
\begin{equation}\label{ilrjakauma_dir_ln}
\hbox{ilr}(\textbf{p}) \approx \mathcal{N} \left( \textbf{V}^T (\hbox{ln}(\boldsymbol{\tilde{\alpha}}_j) - (1/2)((\alpha_K \hbox{exp}(-\mu + \sigma^2/2) + 1)/(\alpha_K + 1)) \  \bm{\lambda} \ \circ \ \boldsymbol{\beta}),   
\gamma(\alpha_K, \mu, \sigma^2) \textbf{V}^T \textbf{D}^{-1}_{\boldsymbol{\tilde{\alpha}}}\textbf{V} \right).
\end{equation}

\section{Simulation study}
In this section,  we  investigate the applicability of the normal approximation (\ref{ilrjakauma_alt}) of the ilr coordinates under different levels of sparsity $\alpha_S$ and the total count $K$. In the following, the sparsity parameter is denoted as $\alpha_S$ 
 instead of $\alpha_K$  as here it will not represent any formal link to the total count $K$. 
In this Section we also investigate how  different levels of variability of the total count influence the performance of the  approximation. 

\subsection{Simulation setup}
We consider four different data generating distributions  (Table \ref{DGMtaulu}): (a) multinomial with fixed total count and fixed class probabilities, (b) Dirichlet-multinomial with variable class probabilities, (c) lognormal-multinomial with variable total count  and (d) lognormal-Dirichlet-multinomial with variable class probabilities and variable total count. 
Table \ref{parametritaulu} presents the values of the model parameters as used in the simulation scenarios.
The number of classes $J = 5$ and the class-specific expected probabilities $\boldsymbol{\tilde{\alpha}}^T = (0.01, 0.04, 0.15, 0.3, 0.5)$ were fixed. 
The SBP matrix was built in a pivotal manner as  in equation (\ref{pivotalmat}), which will allow  investigation of  coordinates  involving both rare and more abundant classes, as well as only abundant classes. We varied the value of $\alpha_S$, which defines the sparseness of the count data. The values for  $\tilde\alpha_j$ and $\alpha_S$ were chosen so that the unimodality condition was fulfilled, i.e., each $\tilde\alpha_j\alpha_S > 1$. 
We also  varied the total count $K$ or, under lognormal counts, its median exp($\mu$) and variability $\sigma^2$. 
The coefficient of variation (CV) of the lognormally distributed total count is  $\sqrt{\hbox{exp}(\sigma^2)-1}$, independent of $\mu$. The values 0.1 and 1.0 for $\sigma^2$ correspond to CV's of 0.32 and 1.31, respectively. In empirical microbiome data, the CV typically falls within this range \citep{aatsinki2018gut,keskitalo2021gut, aatsinki2019gut,aatsinki2020maternal}. 

Table \ref{varianssit} presents the excess variability of the class-specific proportions and,  subsequently of the ilr coordinates   under each data generating mechanism and each parameter setting.  The excess variability was calculated as the ratio of the elements of the variance-covariance matrix of the proportions to those under the multinomial distribution, based on Equations (\ref{excessvar_Dirmn}) and (\ref{dir_ln_mn_exvar}).

Each simulation consisted of 10000 independent draws of counts from their data generating distribution. In case of zero-count observations, occurring in particular under strong sparsity or when the total number of counts was small,
zeroes were replaced with 0.5.  This approach is commonly used in e.g. microbiome studies \citep{Zhang2019a, Sohn2019}.  Based on the 10000 draws, we calculated  the empirical expectations and variance-covariance matrices of the ilr coordinates as approximations to their true values.  

We  investigated the performance of approximation (\ref{ilrjakauma_alt}) by comparing its expectation and eigenvalues of the variance-covariance matrix to their empirical counterparts as found by simulation.  
For each combination of  $\alpha_S$  and $K$, the comparison was made by plotting the logarithmic ratios of the empirical and approximate expectations and eigenvalues.  
The normality of the coordinates  was further assessed by comparing the quantiles of the simulated ilr coordinates to the theoretical quantiles using normal Q-Q-plots. 

\subsection{ Illustration of the composition of  counts and proportions}
\begin{figure}[!htbp]
   \centering
\centerline{\includegraphics[scale=0.6]{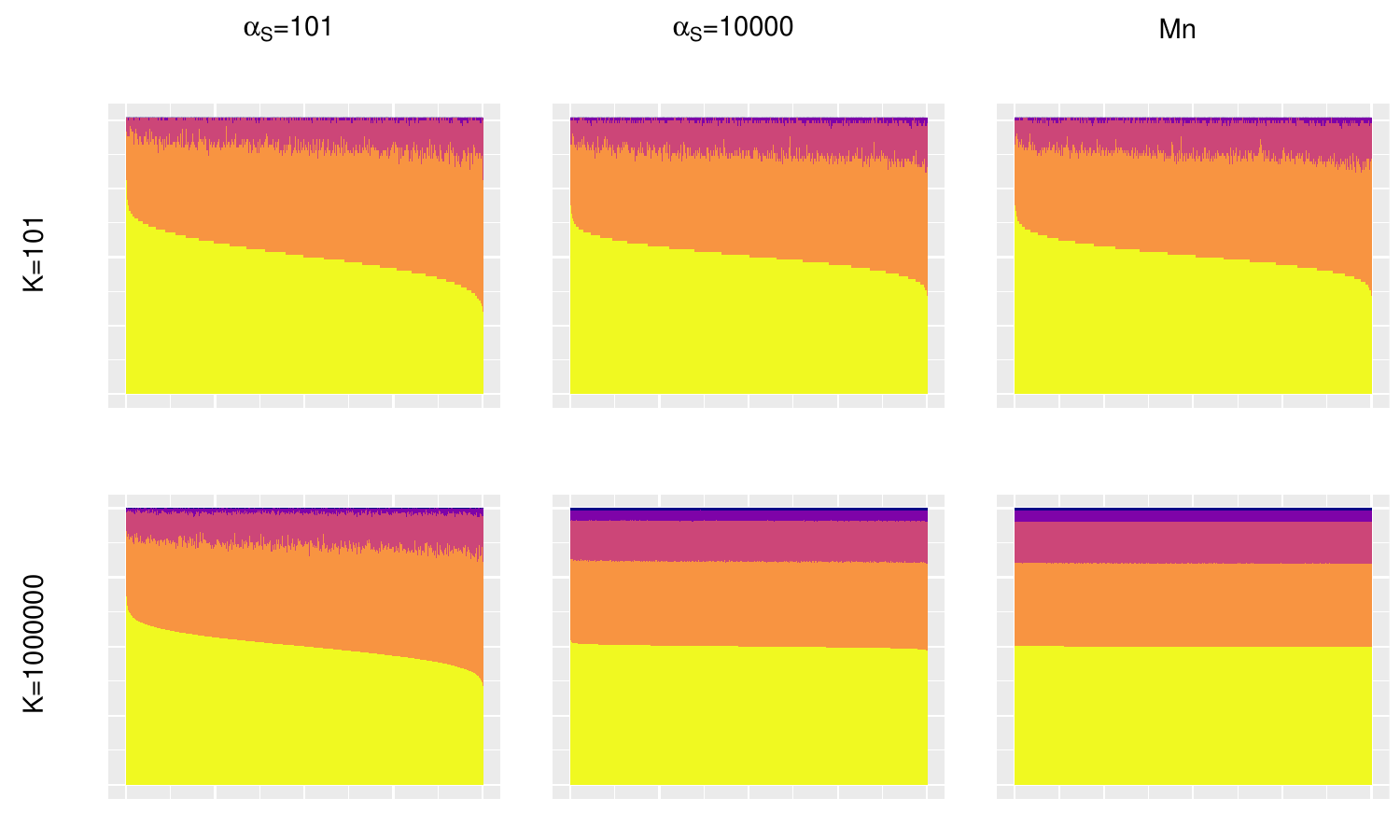}}
    \caption{Compositions of counts under the multinomial (data generating distribution (a); right-hand column) and Dirichlet-multinomial (data generating distribution (b); left-hand and middle column) distributions when the total count $K$ is either 101 (upper panels) or 1000000 (lower panels). Each vertical line represents one observation with the five counts stacked from $K_1$ (upmost) to $K_5$ (lowest) and indicated by the colours corresponding to the five class-specific counts with class probabilities $\tilde{\boldsymbol{\alpha}} = (0.01, 0.04, 0.15, 0.3, 0.5)$.     
    The observations are sorted in descending order of $K_5$. 
As the total count is fixed, the same compositions apply to proportions.  } \label{compcount}
\end{figure}

Before considering the distribution of the ilr coordinates, we illustrate the composition of counts under selected values of $K$ and $\alpha_S$ (Fig.  \ref{compcount}). As the total count $K$ is here fixed, the same compositions apply to  class proportions.  We investigate scenarios with either small ($K =101$) or large ($K=1000000$) total count, and with either high sparsity ($\alpha_S = 101$), intermediate sparsity ($\alpha_S=10000$), or no sparsity (multinomial).  With a large total count, the counts exhibit less  variation than under a small total count.  With increasing value of $\alpha_S$, the distribution of the counts approaches that of multinomial and  the compositions exhibit decreasing amount of heterogeneity. 
 Following the choices of the $\alpha_S$ parameter, the distributions of the counts are unimodal. 
In particular, for each combination of $\alpha_S$ and $K$, the most common proportion is $p_5$ in nearly all samples shown in Fig. 1, and the composition of  the proportions under $\alpha_S=101$ does not largely differ from that under the multinomial distribution.

\subsection{Comparison of the approximate and empirical distributions of the ilr coordinates}\label{validity}
\begin{figure}[!htbp]
    \centering
\includegraphics[scale=0.4]{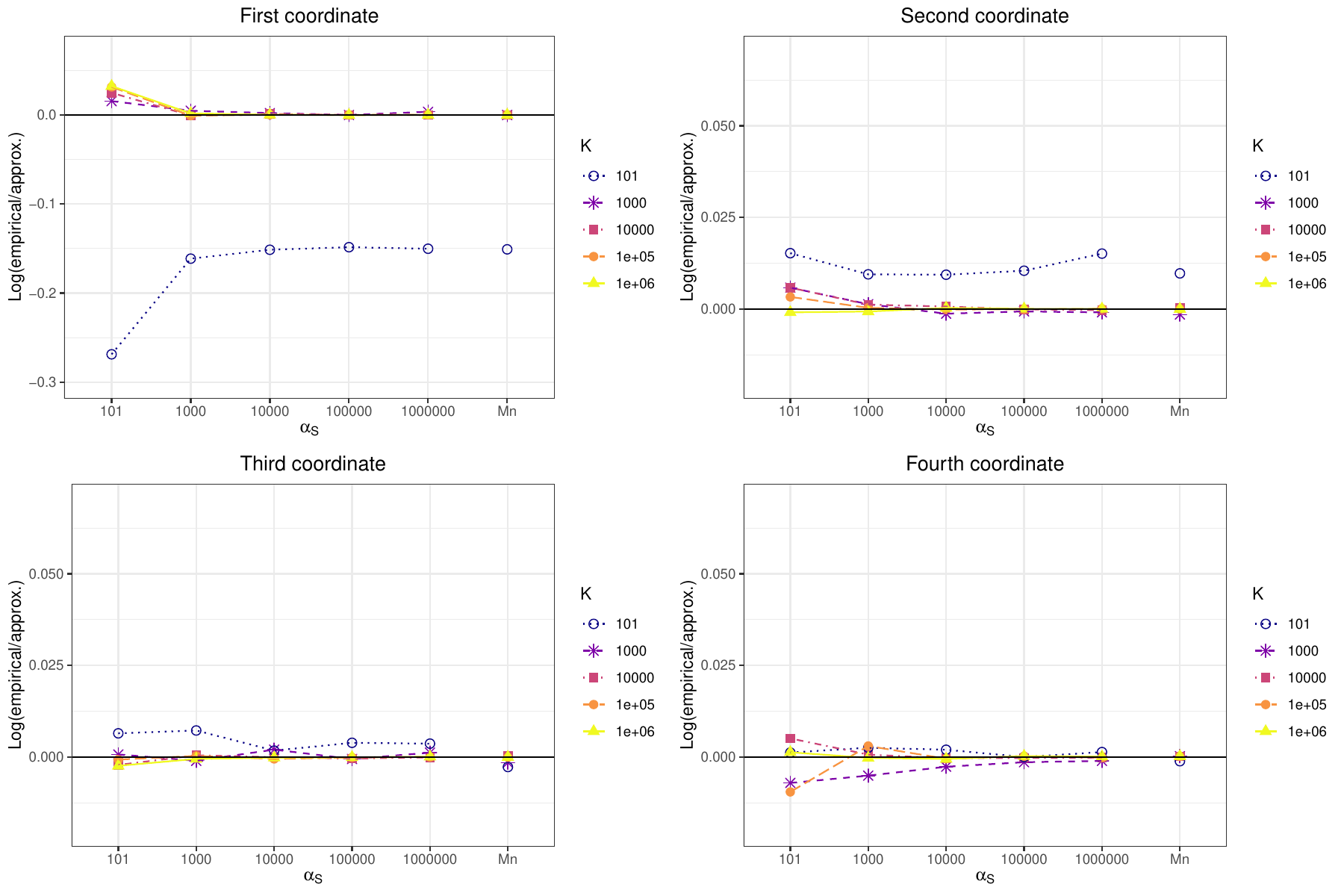}
    \caption{Log-ratios of the empirical expectations based on Monte Carlo simulation and the approximated expected values (Equation \ref{ilrjakauma_alt}) for the four ilr coordinates under multinomial (Mn; data generating distribution (a) in  Table \ref{DGMtaulu}) and Dirichlet-multinomial counts (connected dots; data generating distribution (b) in Table \ref{DGMtaulu}). Value  0  means perfect correspondence between the empirical and approximated values. Note that the scale of the vertical axis for the first coordinate is different from the rest. 
       } \label{ilr_E_vert}
\end{figure}

We here examine the performance of approximation (\ref{ilrjakauma_alt}) through comparisons  with the empirical distribution  of the ilr coordinates.
Under multinomial counts, the approximation of the expected values works well when the total count $K$ is larger than 100  (Fig. \ref{ilr_E_vert},  Mn), corresponding to the findings in Graffelman et al. \citep{Graffelman}.
Also under the Dirichlet-multinomial distribution, the  approximation  holds quite well  when both $\alpha_S \rightarrow \infty$ (i.e., when the distribution of counts approaches multinomial) and  $K \rightarrow \infty$  (Fig. \ref{ilr_E_vert}, connected points). In this case, the bias remains small and is mostly $<  1 \%$.  When either $\alpha_S$ or $K$ is small ($\alpha_S = 101$ or $K=101$), the approximated expectations, especially for the first coordinate,  deviate more from the empirical  value with the absolute bias up to 30 $\%$. 
Interestingly, the approximation appears to be more accurate for the last three ilr coordinates than the first one. The accuracy depends on which proportions are contrasted in each of the coordinates. While the first coordinate involves also the rarest classes, the last three coordinates are only based on the more abundant ones. 

\begin{figure}[!htbp]
    \centering
\includegraphics[scale=0.4]{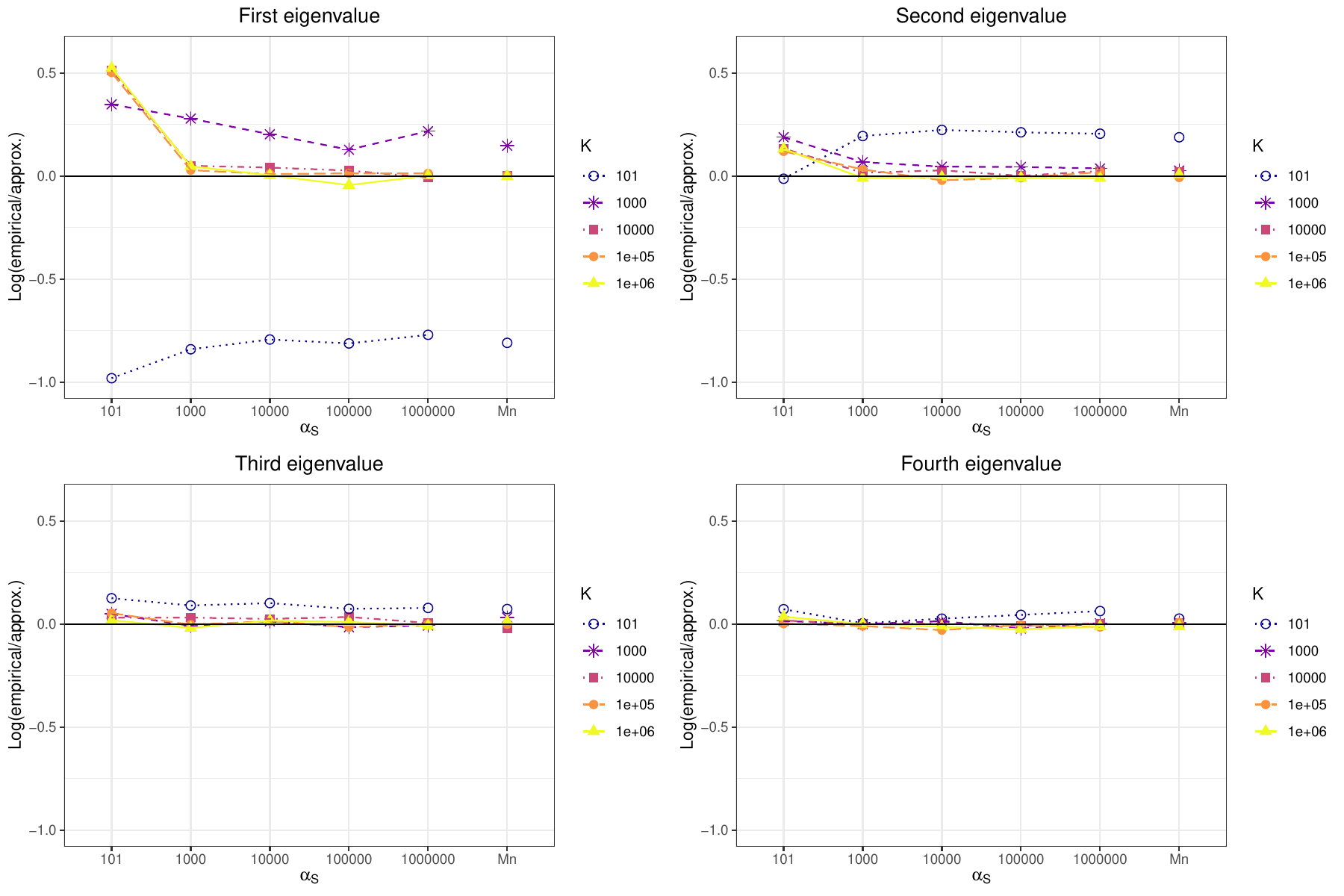}
    \caption{Log-ratios of the eigenvalues of the empirical variance-covariance matrix based on Monte Carlo simulation and the eigenvalues of the approximated variance-covariance matrix (Equation \ref{ilrjakauma_alt}) for the four ilr coordinates under multinomial (Mn; data generating distribution (a) in  Table \ref{DGMtaulu}) and Dirichlet-multinomial counts (connected dots; data generating distribution (b) in  Table \ref{DGMtaulu}). Value 0 means perfect correspondence between the empirical and approximated values. } \label{ilr_var_vert}
\end{figure}

For each eigenvalue of the variance-covariance matrix,  the approximated values correspond to the empirical values when both $\alpha_S $ and $K$ are large  (Fig. \ref{ilr_var_vert}). When either $\alpha_S$ or $K$ is small, however, the approximation does not work as well. In general, the first eigenvalue of the approximated variance-covariance matrix differs the most from the empirical one.   Especially with the smallest $K$, the approximation for the first eigenvalue  differs from the empirical value, even under multinomial counts.

Apart from checking the first two moments of the approximate distribution, we explored the normality of the first and the fourth ilr coordinates by Q-Q plots (Fig. S1, the first three rows).  With a fixed total count, the last ilr coordinate,  consisting only of the two most common classes, tends to be approximately normal with all values of $K$ and $\alpha_S$. The first coordinate, contrasting the rarest class against all other classes, 
 is normally distributed when $K$ is large ($K=1000000$) and there is very little or no sparsity ($\alpha_S = 10000$ or "Mn"). However, with small $\alpha_S$ or small $K$, the approximation does not perform as well,  as already indicated by the problems with the expecation and the variance-covariance matrix.
The same problems can bee seen in the distribution of the  proportions of the first class (Fig. S2).

 To summarise the above findings, we note that although the excess variability with respect to the multinomial distribution in class specific proportions  becomes excessively large with increasing total count $K$  (Table \ref{varianssit}),  the approximations for both the  expected values and eigenvalues perform well.
 By contrast,  the approximation is not as good with small values of $K$, even when the excess variability is relatively small. It can thus be concluded than when $K$ is large, the distribution of the proportions can diverge from the multinomial distribution without compromising the near-normality of the ilr coordinates.  
  On the other hand, when $K$ is small, a much smaller departure  from the multinomial distribution worsens the performance of the approximation. 

\subsection{Normal approximation based on multinomial counts}
We next demonstrate the pitfalls of using a normal approximation that is based on assuming purely multinomial counts  when the underlying counts actually exhibit extra-multinomial variation. 
Letting $\alpha_S\to\infty$ in (\ref{ilrjakauma_alt}), the approximation becomes (see also \citep{Graffelman})
\begin{equation}\label{multinomiapprox}
\hbox{ilr}(\textbf{p}) \approx \mathcal{N}(\textbf{V}^T \hbox{ln}(\textbf{p}),\frac{1}{K} \textbf{V}^T\textbf{D}_{\tilde\alpha}^{-1}\textbf{V}).
\end{equation}
We here present  results only for the Dirichlet-multinomial data generating distribution. However, conclusions were similar when the total count was lognormally distributed instead of being fixed (data not shown). 
  
\begin{figure}[!htbp]
    \centering
\includegraphics[scale=0.4]{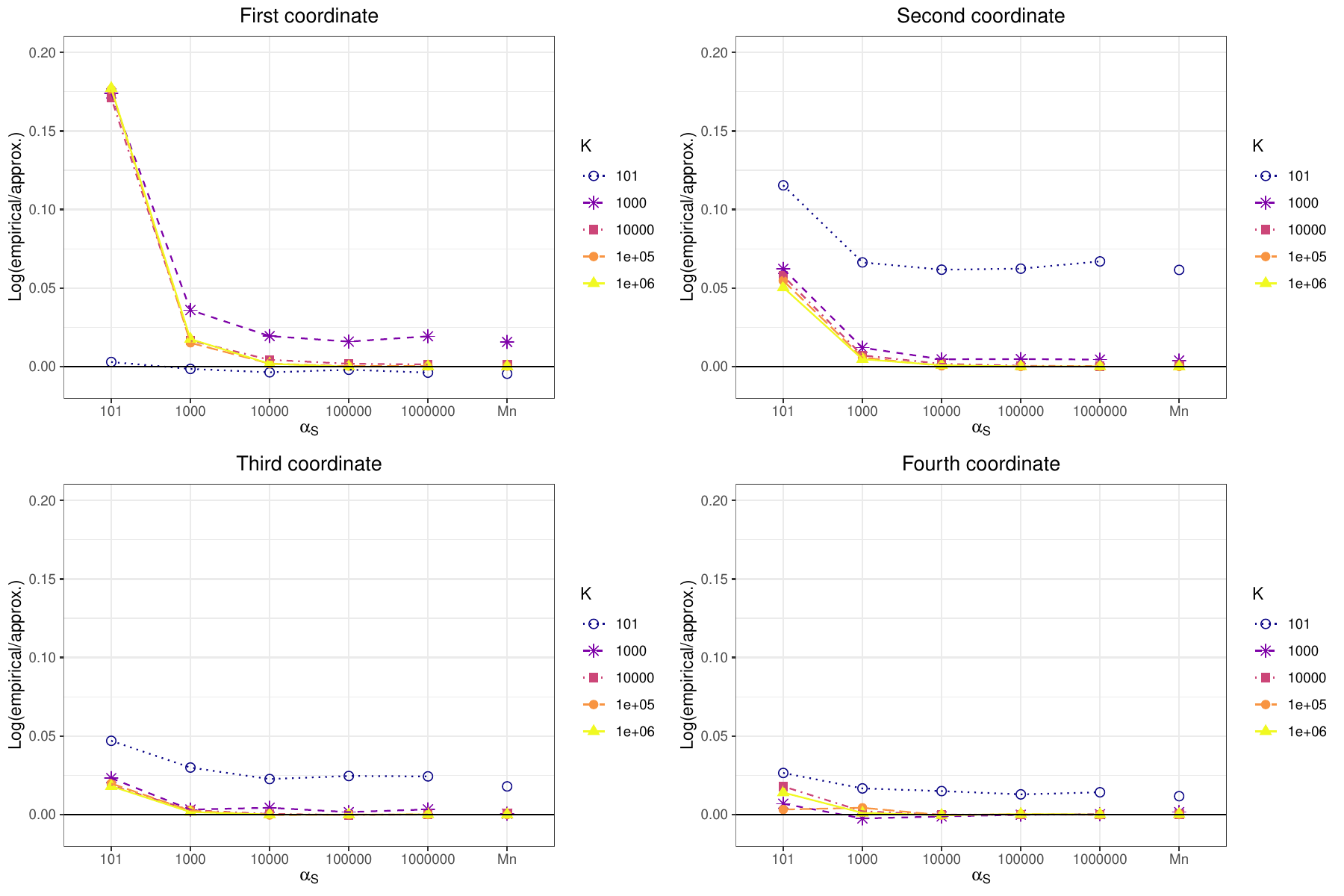}
    \caption{Log-ratios of the empirical expectations based on Monte Carlo simulation and normal approximation  based on multinomial counts under multinomial (Mn) and Dirichlet-multinomial (connected dots) counts. Value  0  means  means perfect correspondence between the empirical and approximated values. Note that the scale differs from that in Figure \ref{ilr_E_vert}. } \label{mn_e}
\end{figure}

Fig. \ref{mn_e}  compares the empirical expectations of the ilr coordinates to those of the multinomial approximation (\ref{multinomiapprox}) in the same simulation scenarios as in Fig. \ref{ilr_E_vert}.  When the counts are sparse (small $\alpha_S$), the  approximation does not hold even under when the total count $K$ is large (bias between empirical and approximated expected values is up to 20 \%).  As expected, when the distribution of the counts approaches multinomial, approximation (\ref{multinomiapprox}) improves.

\begin{figure}[!htbp]
    \centering
\includegraphics[scale=0.4]{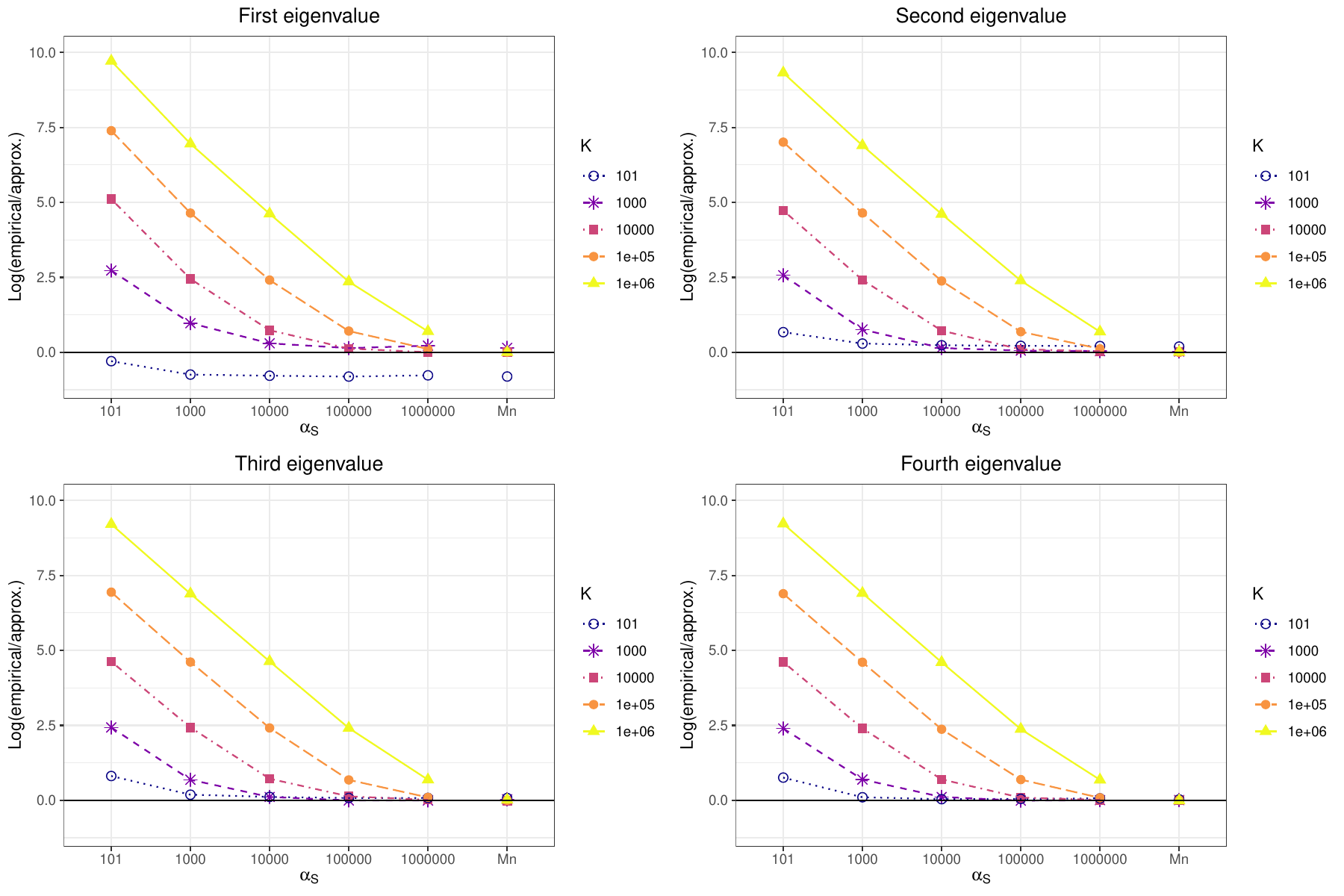}
    \caption{Log-ratios of the eigenvalues of the empirical variance-covariance matrix based on Monte Carlo simulation and eigenvalues of the variance-covariance matrix of the normal approximation based on multinomial counts under multinomial (Mn) and Dirichlet-multinomial (connected dots) counts. Value  0  means  means perfect correspondence between the empirical and approximated values. Note that the scale differs from that in Fig. \ref{ilr_var_vert}.  } \label{mn_var}
\end{figure}

Fig. \ref{mn_var} compares the empirical eigenvalues of the variance-covariance matrix of the ilr coordinates with the corresponding eigenvalues under approximation (\ref{multinomiapprox}) in the same simulation scenarios as in Fig. \ref{ilr_var_vert}. 
The approximate values deviate from the empirical ones when the counts are sparse (small $\alpha_S$). 
However, even under the largest value of  $\alpha_S$, the approximation is not particularly good. 
Somewhat unintuitively, in contrast to the expected values, the performance of the approximation decreases with increasing $K$. 
This  stems from the excess variability in the Dirichlet-multinomial distribution, characterised in Equation (\ref{excessvar_Dirmn}). When $K$ increases, the excess variability also increases.

\subsection{Performance of the approximation in further settings}\label{rare}

\textbf{Variable total count.} We next summarise the performance of approximation (\ref{ilrjakauma_dir_ln}) under lognormally-distributed total count (data generating scenarios (c) and (d)). The variability of the total count ($\sigma^2$)  is evident in the illustration of the composition of the counts (Fig. S3), but after scaling the counts into proportions, the effect of variation in the total count is barely distinguishable (Fig. S4).

The approximation for the expected values  (Fig. S5) performs nearly equally well under overdispersion in the total count as it does under fixed total count (cf. Fig. \ref{ilr_E_vert}). However, when $\mu$ is small, the approximation performs slightly worse for the first two coordinates including also rarer classes. The approximation for the eigenvalues (Fig. S6) of the variance-covariance matrix also works nearly equally well (cf. Fig.  \ref{ilr_var_vert}).  

The distributions of the proportions are quite similar under both fixed and varying total count  (Fig. S2). Importantly, when either $K$ or $\alpha_S$ is small, the distribution of the first proportion is clearly skew under both fixed and variable total count  (Fig. S1). When the expected value of the total count is small or the proportions do not exhibit extra-multinomial variation, the distribution of the fifth, i.e. the most common proportion tends to have heavier tails under variable total count, leading to observations with excess deviation from the mean. This tailedness also follows into the distributions of the ilr coordinates.

\textbf{Rare counts and the choice of contrast matrix.}
We  investigated how the rarity or commonness of the classes and the choice of the contrast matrix affect the performance of approximation (\ref{ilrjakauma_alt}).
First, we investigated a scenario with a pivotal contrast matrix and proportions $\textbf{p} = (0.50, 0.30, 0.15, 0.04, 0.01)$. With this choice, the first ilr coordinate contrasts  the most abundant class against the other classes and  the last ilr coordinate involves  the two rarest classes.  For each coordinate, the approximated expected values now deviate from the empirical values more than with the previous choice of  $\textbf{p} = (0.01, 0.04, 0.15, 0.30, 0.50)$ (Fig. S7) when the total count is small. 
This stems from the rare classes now being included in all ilr coordinates and log-ratio transformation being sensitive to small values. 
The approximation is poorest for the fourth ilr coordinate (bias up to 40 \%).

Second, we investigated a scenario with homogeneous class probabilities $\textbf{p} = (0.20, 0.20, 0.20, 0.20, 0.20)$.  In terms of both expected value and the eigenvalues, the approximation works better than under different class probabilities, even under sparse counts (Fig. S8 and  Fig. S9).

\textbf{Multimodal proportions.}
Finally, we investigated the performance of approximation (\ref{ilrjakauma_alt}) when the distribution of proportions and the ensuing counts is not unimodal, i.e., when some or all of the  Dirichlet parameters $\alpha_j$  are $<1$. 
We set  $\boldsymbol{\alpha} = (0.01, 0.04, 0.15, 0.30, 0.50)$ (with $\alpha_S=1$)  and $\boldsymbol{\alpha} = (0.1, 0.4, 1.5, 3, 5)$ (with  $\alpha_S=10$).
With the first choice,  the multimodality of proportions is evident, as there are samples that nearly exclusively consist of observations in just one class. In general, the compositions of counts exhibit a considerable amount of heterogeneity (Fig. S10). 
The excess sparsity clearly affects the performance of the normal approximation and the empirical expected values and eigenvalues of variance-covariance matrices clearly deviate from the approximated values (Fig. S11 and Fig. S12). Furthermore, especially when all components of $\alpha$ are less than 1, 
the coordinates clearly deviate from the normal distribution (Fig.  S13).

\section{Conclusion\label{sec:4}}
An asymptotic normal approximation to the distribution of the ilr coordinates when based on purely multinomial counts with fixed class probabilities, valid with large enough total count,  has been derived before \citep{Graffelman}. Here, we investigated conditions for the asymptotic normality of ilr coordinates when based on counts with a compound multinomial distribution, i.e., when the multinomial probabilities are not fixed but follow some mixing distribution.  We showed that the asymptotic normality of counts and the resulting ilr coordinates holds when either the mixing distribution converges to a constant or when it admits a suitable limiting distribution. In the special case of a unimodal Dirichlet as the mixing distribution, the asymptotic normality holds when the total count approaches infinity and the variability of the class-specific probabilities across the population goes to zero, irrespective of which of the two converges faster. 

We derived an explicit expression for the normal approximation of the ilr coordinates under Dirichlet-multinomial counts. Our simulation study confirmed the performance of the approximation with large enough total count $K$ and moderate extra-multinomial variability. Importantly, with large enough $K$ the distribution of the proportions can exhibit more extra-multinomial variation without compromising the satisfactory behaviour of the approximation, while under smaller $K$ the distribution of the proportions needs to be closer to multinomial in order to the approximation to perform well. The variability in the class probabilities (regulated by the sparsity parameter $\alpha_K$) affected both the expected value and variance-covariance matrices or the ilr coordinates. When comparing our approximation to an approximation relying on the assumption of multinomial counts, we observed superior performance especially with regard to the eigenvalues of the variance-covariance matrix of the coordinates. Of note, adding variability to the total count did not largely affect the performance of the approximation. Even though in some scenarios it induced heavy tails for the distributions of the coordinates, the performance of the approximation was otherwise good in the scenarios with varying $K$. 

In microbiome data, the read depth determines the total microbial count but is arbitrary and varies across samples. In such situations, the total count is not suitable as the basis for statistical modelling. Instead, the analysis should be based on relative counts whereby by compositional data analysis and models built on ilr coordinates are a viable option \citep{Gloor2017}.  We chose a general multinomial compound distribution to induce overdispersion to the counts and presented conditions under which the resulting ilr coordinates could be modelled using the normal distribution. 

In compositional analysis of microbiome data, the normality of log-ratio transformations is commonly assumed \citep{Sohn2019, Zhang2019a}. However, compositional count observations on the microbiome often exhibit extreme extra-multinomial dispersion (sparsity), i.e. the distribution of counts is not unimodal. Assuming normality of the ilr coordinates may then be ill-justified and, if falsely presumed, may lead to biased or inefficient estimation and false conclusions in mediation analysis as well as other data analytical tasks. In practice, it may be advisable to evaluate the extent of extra-multinomial dispersion in the exploratory phase of the analysis (in the context of Dirichlet distribution, see e.g. \citep{Minka2000}). If statistical modelling is based on assuming normality, taxa which have a non-unimodal distribution should be discarded. A downside of restricting the analysis to taxa whose distribution shows reasonable homogeneity across samples is that potentially interesting information may be lost, if a large number of classes are omitted and the analysis only concerns a sub-part of the original composition. Alternatively, in the context of microbiome data, one may have to investigate only higher taxonomic levels that may be less sparse or to use aggregations, i.e., add up specific parts of composition. Such aggregations can be based on e.g. taxonomic knowledge or defined in a data-driven manner \citep{GordonRodriguez2021}. Based on our results, even under unimodal counts it may be meaningful to build the contrasts in such manner that the rarest classes are not involved in all of the coordinates, given that the empirical research question allows this choice.

Microbiome data are often very high-dimensional, while we here have investigated the distribution of the ilr coordinates in low dimensions only. Nevertheless, we surmise that the asymptotic normality as shown here holds regardless of the dimension, even when the total count is variable. Of note, our treatment of the normal approximation does not depend on the choice of the contrast matrix. Thus, even though we only considered a pivotal SBP matrix in the simulation study, it is conceivable that our results hold also under different matrices. In practice, however, the rarity of specific classes and their position in the contrast matrix, i.e., their role in the coordinates, may affect the performance of the approximation.

The use of the Dirichlet distribution to induce overdispersion in multivariate count data has been criticised due to the negative correlation it creates between the components and the strong independence structure of ratios between different components \citep{ComasCufi2020}. As an alternative, the logratio-multinomial-normal distribution has been proposed in model selection and longitudinal microbiome dynamics \citep{Xia2013, ComasCufi2020, Silverman2018}. Also this approach is a compound multinomial distribution but the mixing distribution is obtained as the inverse transformation of ilr coordinates, which are assumed to follow a normal distribution. Importantly, our aim was not to develop a new statistical model for overdispersed count data under constraints on the total count. Rather, our objective was to address a situation where the total count is not informative and as such is unsuitable for statistical modelling. The motivation to use the Dirichlet-multinomial in the simulation study was due to its transparency and the ease of interpretation under this model.

In addition to sequencing data, such as that on the microbiome, the framework of compositional data analysis is applicable in such research questions where the data arise as non-negative observations that can be scaled by their total \citep{pawlowskyglahn_compositional_2011}. One recent example within the field of health sciences considers partition of daily time use \citep{Pasanen2022}. Another example of measurements that could be considered as compositions is dietary intake of macro-nutrients when characterised as percentage of total energy (E \%). In addition, compositional data analysis and our findings may be applicable in research questions in geology (e.g. mineral composition in a sample) or social sciences (e.g. election polls). Compared to data on the microbiome, compositions based on these kind of data often exhibit less sparsity and less variability in the total count and  the  the normal approximation for the ilr coordinates may thus perform adequately more often than for the microbiome.

\newpage
\section*{Tables}

\begin{table}[!htbp]
\caption{Four data generating distributions and their parameters.}
\renewcommand\arraystretch{2.5} 
\centering
\begin{tabular}{ll | c | c |}
&\multicolumn{1}{c}{}&\multicolumn{2}{c}{\textbf{Total count K}}\\[-2ex]
&\multicolumn{1}{c}{}
&\multicolumn{1}{c}{Fixed}&\multicolumn{1}{c}{Lognormal}\\
\cline{3-4}
\multirow{2}{*}{\rotatebox{90}{\textbf{Prob $\boldsymbol{\Pi}$}}}
&Fixed &a) Multinomial($K$, $\boldsymbol{\Pi}$)&c) Lognormal-multinomial($\mu$, $\sigma^2$, $\boldsymbol{\Pi}$)\\
\cline{3-4}
& Dirichlet &b) Dir-multinomial($K$, $\boldsymbol{\alpha}$)&d) Lognormal-Dir-multinomial($\mu$, $\sigma^2$, $\boldsymbol{\alpha}$)
\\
\cline{3-4}
\end{tabular}
\qquad
\label{DGMtaulu}
\end{table}

\newpage

\begin{table}[!htbp]
\caption{Parameters of the simulation study. 
The sparsity parameter $\alpha_S$ only applies when $\boldsymbol{\Pi}\sim$ Dirichlet. Parameter $\sigma^2$ only applies when $K\sim$ lognormal. The total count is either fixed at $K$ or has a lognormal distribution with median $\mu$ = ln($K$). 
 The four different data generating distributions are indicated by a--d (see Table 1).  }
\centering
\begin{tabular}{l l l }
\textbf{Parm} & \textbf{Values} & \textbf{Note} \\
\hline
$J$ & 5 & Number of classes \\
$\boldsymbol{\Pi}$; $\boldsymbol{\tilde{\alpha}}$ & (0.01, 0.04, 0.15, 0.3, 0.5)  & Class-specific expected probabilities  \\
$\alpha_S$ & 101; 1000; 10000; 100000; 1000000  & Sparsity parameter  (b,d) \\
$K$; $\hbox{exp}(\mu)$  & 101; 1000; 10000; 100000; 1000000 & Total count (a,b); median total count (c,d) \\
$\sigma^2$   & 0.1; 1.0  & Variability of logarithm of total count (c, d) \\ 
\hline
\end{tabular}
\label{parametritaulu}
\end{table}

\newpage

\begin{table}[!htbp]
\caption{Excess variability in class-specific proportions ($p_j$) under some selected simulation scenarios with varying values of sparsity parameter $\alpha_S$ and total count $K$ or, in case of lognormally distributed total count, its median exp($\mu$) and variability $\sigma^2$. The variance-covariance matrix elements are compared against those of multinomial proportions, based on equations (\ref{excessvar_Dirmn}) or (\ref{dir_ln_mn_exvar}).  Abbreviations: DGD=data generating distribution (see Table \ref{DGMtaulu}); Mn=multinomial; Dir=Dirichlet; LN=lognormal.   }
\centering
\begin{tabular}{l l lllllll }
& & & & \multicolumn{5}{c}{$K$} \\ \cline{5-9}
DGD	&	Eq.	&	$\alpha_S$	&	$\sigma^2$	&	101	&	1000	&	10000	&	100000	&	1000000	\\
\hline
(a) Mn	&	-	&	-	&	-	&	1	&	1	&	1	&	1	&	1	\\
(b) Dir-Mn	&	(\ref{excessvar_Dirmn})	&	101	&	-	&	1.98	&	10.79	&	99.03	&	981.38	&	9804.91	\\
(b) Dir-Mn	&	(\ref{excessvar_Dirmn})	&	1000	&	-	&	1.1	&	2	&	10.99	&	100.9	&	1000	\\
(b) Dir-Mn	&	(\ref{excessvar_Dirmn})	&	10000	&	-	&	1.01	&	1.1	&	2	&	11	&	100.99	\\
(b) Dir-Mn	&	(\ref{excessvar_Dirmn})	&	100000	&	-	&	1	&	1.01	&	1.1	&	2	&	11	\\
(b) Dir-Mn	&	(\ref{excessvar_Dirmn})	&	1000000	&	-	&	1	&	1	&	1.01	&	1.1	&	2	\\
(c) LN-Mn	&	(\ref{dir_ln_mn_exvar})	&	-	&	0.1	&	1.05	&	1.05	&	1.05	&	1.05	&	1.05	\\
(d) LN-Dir-Mn	&	(\ref{dir_ln_mn_exvar})	&	101	&	0.1	&	2.03	&	10.84	&	99.08	&	981.43	&	9804.96	\\
(d) LN-Dir-Mn	&	(\ref{dir_ln_mn_exvar})	&	1000	&	0.1	&	1.15	&	2.05	&	11.04	&	100.95	&	1000.05	\\
(d) LN-Dir-Mn	&	(\ref{dir_ln_mn_exvar})	&	10000	&	0.1	&	1.06	&	1.15	&	2.05	&	11.05	&	101.04	\\
(d) LN-Dir-Mn	&	(\ref{dir_ln_mn_exvar})	&	100000	&	0.1	&	1.05	&	1.06	&	1.15	&	2.05	&	11.05	\\
(d) LN-Dir-Mn	&	(\ref{dir_ln_mn_exvar})	&	1000000	&	0.1	&	1.05	&	1.05	&	1.06	&	1.15	&	2.05	\\
(c) LN-Mn	&	(\ref{dir_ln_mn_exvar})	&	-	&	1	&	1.65	&	1.65	&	1.65	&	1.65	&	1.65	\\
(d) LN-Dir-Mn	&	(\ref{dir_ln_mn_exvar})	&	101	&	1	&	2.62	&	11.44	&	99.67	&	982.02	&	9805.55	\\
(d) LN-Dir-Mn	&	(\ref{dir_ln_mn_exvar})	&	1000	&	1	&	1.75	&	2.65	&	11.64	&	101.55	&	1000.65	\\
(d) LN-Dir-Mn	&	(\ref{dir_ln_mn_exvar})	&	10000	&	1	&	1.66	&	1.75	&	2.65	&	11.65	&	101.64	\\
(d) LN-Dir-Mn	&	(\ref{dir_ln_mn_exvar})	&	100000	&	1	&	1.65	&	1.66	&	1.75	&	2.65	&	11.65	\\
(d) LN-Dir-Mn	&	(\ref{dir_ln_mn_exvar})	&	1000000	&	1	&	1.65	&	1.65	&	1.66	&	1.75	&	2.65	\\
\hline
\end{tabular}
\label{varianssit}
\end{table}

\newpage

\section*{Figures}

\newpage

\bibliographystyle{plainnat}
\bibliography{References}

\section*{Financial disclosure}
There are no financial conflicts of interest to disclose. 
\section*{Conflict of interest}
The authors declare no conflicts of interest. 
\section*{Author contributions}
NK and KA conceived the presented idea and wrote the manuscript. NK performed the the empirical simulation study. 
JV developed the theoretical proof presented in the supplementary material. JN critically reviewed the paper and contributed to the interpretation of the results. 
KA, JN and OR supervised the manuscript. All authors discussed the research and provided critical feedback that helped shape the manuscript.

\section*{Appendix}

\textbf{Notation.} Denote the $J$-dimensional closed unit simplex by $\mathcal{S}^{J - 1}$. Let $X_K \mid \Pi_K \sim \mathrm{Multinomial}(K, \Pi_K)$, where the random vector $\Pi_K$ satisfies $\Pi_K \in \mathcal{S}^{J - 1}$ and 
\[ 
\Pi_K = (\Pi_{K1}, \ldots , \Pi_{K J})' = {\bm\alpha} + Z_K,
\]
for some fixed ${\bm\alpha} \in \mathcal{S}^{J - 1}$ and for some sequence of random $J$-vectors $Z_K = (Z_{K1}, \ldots, Z_{KJ})'$.
Let $\hbox{diag}(\Pi_K)$ denote a diagonal matrix with elements of $\Pi_K$ on the diagonal. Any $h \in \mathcal{S}^{J - 1}$ induces a partitioning of $[0, 1]$ into the intervals $[0, h_1], (h_1, h_1 + h_2], \ldots,  (\sum_{\ell=1}^{J - 1} h_\ell, 1]$. We denote the $j$th interval in this partitioning by $\mathcal{P}_j(h)$. Finally, we denote the Lebesgue measure of a subset $\mathcal{A}$ of the real line by $\lambda(\mathcal{A})$.

\begin{proof}[\textbf{\upshape Proof of Theorem 1:}]
	
We first observe that $X_K$, conditionally on $\Pi_K$, has the same distribution as the vector $T_K = (T_{K1}, \ldots, T_{KJ})'$, where
\begin{align*}
T_{Kj} := \sum_{k = 1}^K \mathbb{I} \left( U_k \in \mathcal{P}_j(\Pi_K) \right)
\end{align*}
and $U_k$ are i.i.d. $U(0, 1)$-variates independent of $\Pi_K$ and $\mathbb{I}$ denotes the indicator function. The indicators above then satisfy,
\begin{align*}
& \mathbb{I}(U_k \in \mathcal{P}_j(\Pi_K)) - \mathbb{I}(U_k \in \mathcal{P}_j({\bm\alpha})) \\
=& \mathbb{I}(U_k \in \mathcal{P}_j(\Pi_K) \setminus \mathcal{P}_j({\bm\alpha}))  - \mathbb{I}(U_k \in \mathcal{P}_j({\bm\alpha}) \setminus \mathcal{P}_j(\Pi_K)).
\end{align*}
Thus the $j$th component of $T_K$ satisfies 
\begin{align}\label{eq:main_thing}
\begin{split}
\frac{1}{\sqrt{K}}(T_{Kj} - K \alpha_j)  =& \frac{1}{\sqrt{K}} \left( \sum_{k = 1}^K \mathbb{I}(U_k \in \mathcal{P}_j({\bm\alpha})) - K \alpha_j \right)  \\
+& \frac{1}{\sqrt{K}}  \sum_{k = 1}^K \mathbb{I}(U_k \in \mathcal{P}_j(\Pi_K) \setminus \mathcal{P}_j({\bm\alpha})) \\
-& \frac{1}{\sqrt{K}} \sum_{k = 1}^K \mathbb{I}(U_k \in \mathcal{P}_j({\bm\alpha}) \setminus \mathcal{P}_j(\Pi_K)) .
\end{split}
\end{align}
We next show that the second term on the RHS of \eqref{eq:main_thing} is negligible in probability. Denoting the second term by $S_K$, we observe that, conditional on $Z_K$,
\begin{align*}
\sqrt{K} S_K  \mid Z_K \sim \mathrm{Bin}(K, \lambda(\mathcal{P}_j(\Pi_K) \setminus \mathcal{P}_j({\bm\alpha}))).
\end{align*}
Noting further that
\begin{align*}
\lambda(\mathcal{P}_j(\Pi_K) \setminus \mathcal{P}_j({\bm\alpha})) \leq \left| \sum_{\ell = 0}^{j - 1} \Pi_{K \ell} - \sum_{\ell = 0}^{j - 1} \alpha_{\ell} \right| + \left| \sum_{\ell = 0}^{j} \Pi_{K \ell} - \sum_{\ell = 0}^{j} \alpha_{\ell} \right|  \leq 2 \sqrt{J} \| Z_K \|,
\end{align*}
where we define $\Pi_{K 0} = \alpha_{0} := 0$, we get
\begin{align*}
\mathrm{E}(S_K) = \frac{1}{\sqrt{K}} \mathrm{E}\{ \mathrm{E}( \sqrt{K} S_K \mid Z_K) \}) \leq 2 \sqrt{KJ} \mathrm{E}\| Z_K \|,
\end{align*}
which is of the order $o(1)$ by our assumptions since $\{ \sqrt{K} \mathrm{E}\|Z_K\|\}^2 \leq K \mathrm{E} ( \|Z_K\|^2 ) $. Similarly, we get for the variance that
\begin{align*}
\mathrm{Var}(S_K) =& K \mathrm{Var}(\lambda_K) + \mathrm{E}\{ \lambda_K (1 - \lambda_K) \} \\
=& (K - 1) \mathrm{E}(\lambda_K^2) - K \{ \mathrm{E}(\lambda_K)\}^2 + \mathrm{E}(\lambda_K).
\end{align*}
where $\lambda_K := \lambda(\mathcal{P}_j(\Pi_K) \setminus \mathcal{P}_j(\boldsymbol\alpha))$. Arguing as before, we see that the last two terms above are of the order $o(1)$. For the first term, we have
\begin{align*}
(K - 1) \mathrm{E}(\lambda_K^2) \leq 4 K J \mathrm{E}\| Z_K \|^2,
\end{align*}
which is $o(1)$ by our assumptions.

Thus the second term on the RHS of \eqref{eq:main_thing} is negligible in probability. Similarly one can show that also the third term is of the order $o_p(1)$. Thus the joint limiting distribution of the elements of $T_K$ is determined solely by the respective first terms on the RHS of \eqref{eq:main_thing}. The claim now follows from the standard central limit theorem.
\end{proof}

\begin{proof}[\textbf{\upshape Proof of Theorem 2:}]
	We first decompose as follows,
	\begin{align*}
	\frac{\sqrt{\alpha_K}}{K}(X_K - K {\bm\alpha}) = \frac{\sqrt{\alpha_K}}{K}(X_K - K \Pi_K) + \sqrt{\alpha_K}(\Pi_K - {\bm\alpha}),
	\end{align*}
	where the second term has the limiting distribution $\mathcal{D}$. Hence, it remains to show that the first term is negligible in probability. We achieve this by establishing that its first two moments vanish when $K \rightarrow \infty$.
	
	By the law of total expectation (conditioning on $\Pi_K$),
	\begin{align*}
	\frac{\sqrt{\alpha_K}}{K}\mathrm{E}(X_K - K \Pi_K) = \frac{\sqrt{\alpha_K}}{K} \mathrm{E}(K \Pi_K - K \Pi_K) = 0.
	\end{align*}
	Similarly, by the law of total covariance,
	\begin{align*}
	& \mathrm{Cov} \left\{ \frac{\sqrt{\alpha_K}}{K}(X_K - K \Pi_K) \right\}\\
	=& \frac{\alpha_K}{K^2} \left[ \mathrm{Cov}\{ \mathrm{E} ( X_K - K \Pi_K \mid \Pi_K) \} + \mathrm{E}\{ \mathrm{Cov} ( X_K - K \Pi_K \mid \Pi_K) \} \right] \\
	=& \frac{\alpha_K}{K^2} K \mathrm{E}\{ \hbox{diag}(\Pi_K)
  - \Pi_K \Pi_K' \}.
	\end{align*}
	The resulting expectation term is bounded as can be seen by writing,
	\begin{align*}
	 \| \mathrm{E}\{\hbox{diag}(\Pi_K) - \Pi_K \Pi_K' \} \|_1 
	\leq& \mathrm{E}\| \hbox{diag}(\Pi_K) - \Pi_K \Pi_K' \|_1 \\
	\leq& \mathrm{E}\| \hbox{diag}(\Pi_K) \|_1 + \mathrm{E}\| \Pi_K \Pi_K' \|_1 \\
	\leq& J + J^2,
	\end{align*}
	where $\| \cdot \|_1$ is the element-wise $\ell_1$-norm. The claim now follows.
\end{proof}

\begin{proof}[\textbf{\upshape Proof of Corollary 1:}]
The expectation of $\Pi_K$ is simply $\tilde{\bm\alpha}$ and its covariance matrix is
	\begin{align*}
	\frac{1}{\alpha_K + 1} \{  \hbox{diag}(\boldsymbol{\tilde\alpha}) - \tilde{\bm\alpha} \tilde{\bm\alpha}' \}.
	\end{align*}
	We then have for $Z_K = \Pi_K - \tilde{\bm\alpha}$ that
	\begin{align*}
	\mathrm{E} ( \|Z_K\|^2 ) =& \mathrm{E}\{ \Pi_K - \mathrm{E}(\Pi_K) \}' \{ \Pi_K - \mathrm{E}(\Pi_K) \} 
	= \mathrm{tr}\{ \mathrm{Cov}(\Pi_K) \} 
	= \frac{1}{\alpha_K + 1} (1 - \| \tilde{\bm\alpha} \|^2),
	\end{align*}
	from which the claim now follows by invoking Theorem \ref{theorem_1}.
\end{proof}	

\begin{proof}[\textbf{\upshape Proof of Corollary 2:}]
 By Theorems 4.2 and 4.3 in Geyer and Meeden (2013), the random vector $\Pi_K$ satisfies,
	\begin{align*}
	\sqrt{\alpha_K}\left( \Pi_K - \frac{\alpha_K \tilde{\bm\alpha} - 1_J}{\alpha_K - J}\right) \rightsquigarrow \mathcal{N}(0, \hbox{diag}(\boldsymbol{\tilde\alpha}) - \tilde{\bm\alpha} 
\tilde{\bm\alpha}'),
	\end{align*}
	where $1_J$ is a vector of ones. This implies that
	\begin{align*}
	 \sqrt{\alpha_K+1}(\Pi_K - \tilde{\bm\alpha}) 
	=&\sqrt{\frac{\alpha_K+1}{\alpha_K}} 	\left( \sqrt{\alpha_K}\left( \Pi_K - \frac{\alpha_K \tilde{\bm\alpha} - 1_J}{\alpha_K - J}\right) + \sqrt{\alpha_K}\left( \frac{\alpha_K \tilde{\bm\alpha} - 1_J}{\alpha_K - J} - \tilde{\bm\alpha} \right) \right) \\
	=& \sqrt{\frac{\alpha_K+1}{\alpha_K}} 	\left(  \sqrt{\alpha_K}\left( \Pi_K - \frac{\alpha_K \tilde{\bm\alpha} - 1_J}{\alpha_K - J}\right) + \sqrt{\alpha_K} \frac{J \tilde{\bm\alpha} - 1_J}{\alpha_K - J} \right) \\
	& \rightsquigarrow \mathcal{N}(0,\hbox{diag}(\boldsymbol{\tilde\alpha}) - \tilde{\bm\alpha} \tilde{\bm\alpha}').
	\end{align*} 
	The claim now follows by invoking Theorem \ref{theorem_2}.
\end{proof}
\newpage

\section*{Supporting information}
The following supporting information is available as a supplementary material. 

 \begin{itemize} 
  \item[]  Figure S1. Quantile-quantile plot for the first (ilr$_1$) and fourth (ilr$_4$) ilr coordinate under different parameter combinations. The x axis corresponds to theoretical and y axis to sample quantiles.  The notation $\sigma^2$ = NA refers to the Dirichlet-multinomial scenario with fixed total count.
 \item[] Figure S2. Quantile-quantile plot for the first (p$_1$) and last (p$_5$) proportion under different parameter combinations. The x axis corresponds to theoretical and y axis to sample quantiles. The notation $\sigma^2$ = NA refers to the Dirichlet-multinomial scenario with fixed total count.
 \item[] Figure S3. Compositions of  counts under lognormal-multinomial (data generating distribution (c); right-hand column) and lognormal-Dirichlet-multinomial (data generating distribution  (d); left-hand and middle column) counts when the $\mu$ is either 101 or 1000000. Each vertical line represents one observation with the five counts stacked from $K_1$ (upmost) to $K_5$ (lowest) and indicated by the colours. The observations are sorted in descending order of $K_5$. 
 \item[] Figure S4. Compositions of  proportions under lognormal-multinomial (data generating distribution (c); right-hand column) and lognormal-Dirichlet-multinomial (data generating distribution  (d); left-hand and middle column) counts when the $\mu$ is either 101 or 1000000.  Each vertical line represents one observation with the five proportions stacked from $p_1$ (upmost) to $p_5$ (lowest) and indicated by the colours. The observations are sorted in descending order of $p_5$. 
 \item[] Figure S5. Log-ratios of the empirical expectations based on Monte Carlo simulation and the approximated expected values of the four ilr coordinates under lognormal-multinomial (Mn; data generating distribution (c)) (see Table 1 in the main text) and lognormal-Dirichlet-multinomial counts (connected dots; data generating distribution (d)). Value  0  means  means perfect correspondence between the empirical and approximated values. Note that the scale of the vertical axis for the first coordinate is different from the rest. 
 \item[] Figure S6. Log-ratios of the eigenvalues of the empirical variance-covariance matrix based on Monte Carlo simulation and the eigenvalues of the approximated variance-covariance matrix for the four ilr coordinates under lognormal-multinomial (Mn; data generating distribution (c)) (see Table 1 in the main text) and lognormal-Dirichlet-multinomial counts (connected dots; data generating distribution (d)). Value 0 means perfect correspondence between the empirical and approximated values.
 \item[] Figure S7. Log-ratios of the empirical expectations based on Monte Carlo simulation and the approximated expected values of the four ilr coordinates under multinomial (Mn; data generating distribution (a)) (see Table 1 in the main text) and Dirichlet-multinomial counts (connected dots; data generating distribution (b)), when the order of proportions used in the ilr transformation is $\boldsymbol{\tilde\alpha} = (0.50, 0.30, 0.15, 0.04, 0.01)$. Value  0  means  means perfect correspondence between the empirical and approximated values. 
 \item[] FigureS 8. Differences of the empirical expectations based on Monte Carlo simulation and the approximated expected values of the four ilr coordinates under multinomial (Mn; data generating distribution (a)) (see Table 1 in the main text) and Dirichlet-multinomial counts (connected dots; data generating distribution (b)), when the order of proportions used in the ilr transformation is $\boldsymbol{\tilde\alpha} = (0.2, 0.2, 0.2, 0.2, 0.2)$. Value 0means perfect correspondence between the empirical and approximated values. 
 \item[] Figure S9. Log-ratios of the eigenvalues of the empirical variance-covariance matrix based on Monte Carlo simulation and the eigenvalues of the approximated variance-covariance matrix for the four ilr coordinates under multinomial (Mn; data generating distribution (a)) (see Table 1 in the main text) and Dirichlet-multinomial counts (connected dots; data generating distribution (b)), when the order of proportions used in the ilr transformation is $\boldsymbol{\tilde\alpha} = (0.2, 0.2, 0.2, 0.2, 0.2)$.  Value 0 means perfect correspondence between the empirical and approximated values. 
 \item[] Figure S10. Compositions of  proportions under selected scenarios when the distribution of the proportions is multimodal ($\alpha_S = 1$; $\alpha_S = 10$).  Each vertical line represents one observation with the five counts stacked from $p_1$ (upmost) to $p_5$ (lowest) and indicated by the colours. The observations are sorted in descending order of $K_5$. 
 \item[] FigureS11. Log-ratios of the empirical expectations based on Monte Carlo simulation and the approximated expected values of the four ilr coordinates under multinomial (Mn; data generating distribution (a)) (see Table 1 in the main text) and Dirichlet-multinomial counts (connected dots; data generating distribution (b)), including the multimodal scenarios where $\alpha_S = 1$ and $\alpha_S = 10$. Value  0  means  means perfect correspondence between the empirical and approximated values.  Note that the scale of the vertical axis for the first and second coordinates is different from the rest. 
 \item[] Figure S12. Log-ratios of the eigenvalues of the empirical variance-covariance matrix based on Monte Carlo simulation and the eigenvalues of the approximated variance-covariance matrix for the four ilr coordinates under multinomial (Mn; data generating distribution (a)) (see Table 1 in the main text) and Dirichlet-multinomial counts (connected dots; data generating distribution (b)), including the multimodal scenarios where $\alpha_S = 1$ and $\alpha_S = 10$. Value 0 means perfect correspondence between the empirical and approximated values.
 \item[] Figure S13 .Quantile-quantile plot for the first (ilr$_1$) and fourth (ilr$_4$) ilr coordinate under different parameter combinations when the distribution of the proportions is multimodal ($\alpha_S = 1$; $\alpha_S = 10$).  The x axis corresponds to theoretical and y axis to sample quantiles.  
 \end{itemize}

\end{document}